\documentclass[twocolumn,aps,showpacs]{revtex4}
\usepackage{amssymb}
\usepackage{epsf}
\usepackage{amsmath}
\usepackage{graphicx}

\def\lsim{\lower -0.3ex \hbox{$<$} \kern -0.75em \lower 0.7ex \hbox{$\sim$}}
\def\gsim{\lower -0.3ex \hbox{$>$} \kern -0.75em \lower 0.7ex \hbox{$\sim$}}
\def\Journal #1,#2,#3,#4#5#6#7{#1 {\bf #2}, #3 (#4#5#6#7)}

\newcommand{\GVec}[1]{\mbox{\boldmath$#1$}}

\def\Vec#1{{\bf #1}}
\def\GVec#1{\mbox{\boldmath $#1$}}

\def\vare{\varepsilon}

\def\partd#1#2{\frac{\partial #1}{\partial #2}}

\begin{document}

\title{
Electronic transmission through
AB-BA domain boundary in bilayer graphene
}
\author{Mikito Koshino}
\affiliation{
Department of Physics, Tohoku University, Sendai, 980-8578, Japan
}

\begin{abstract}
We study the electron transmission through
the domain boundary on bilayer graphene
separating AB and BA stacking regions.
Using the effective continuum model,
we calculate the electron transmission probability 
as a function of the electron energy and the incident angle,
for several specific boundary structures.
The transmission strongly depends on
the crystallographic direction of the boundary
and also on the atomic configuration inside.
At the low energy, the boundary is either insulating or 
highly transparent depending on the structure.
In insulating cases, the transmission
sharply rises when the Fermi energy is increased to a certain level,
suggesting that the electric current through the boundary 
can be controlled by the field effect.
The boundary parallel to the zigzag direction 
generally have different transmission properties
between the two different valleys, and this enables
to generate the valley polarized current
in a certain configuration.
We show that those characteristic features can be
qualitatively explained by the 
transverse momentum conservation in
the position-dependent band structure
in the intermediate region.
\end{abstract}


\maketitle

\section{Introduction}

\begin{figure}
\centerline{\epsfxsize=0.8\hsize \epsffile{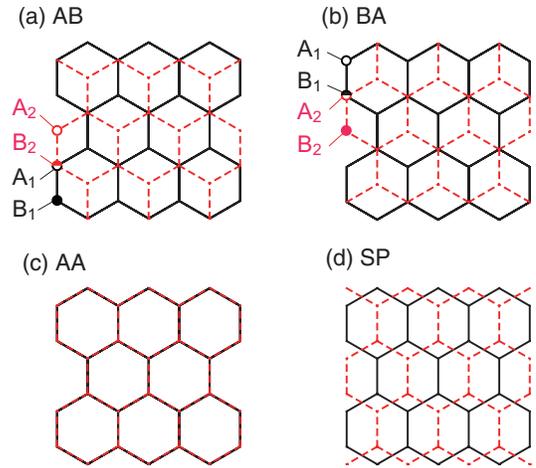}}
\caption{
Various stacking structures of bilayer graphene.
(a) AB,  (b) BA, (c) AA, and (d) SP stacking.
}
\label{fig_stacking}
\end{figure}

In multilayer graphenes, the weak interlayer interaction 
allows various different stacking configurations, 
and there often appear
the domain structures consisting of different stacking regions.
\cite{lui2010imaging,ping2012layer,brown2012twinning,hattendorf2013networks,alden2013strain}
Recent experiment found a novel type of domain boundary
in bilayer graphene \cite{brown2012twinning,alden2013strain} 
which connects two equivalent, but distinct Bernal-stacking structures 
referred to AB and BA, which are shown in Fig.\ \ref{fig_stacking}(a) and (b),
respectively.
On the boundary, the AB structure is continuously deformed to
the BA structure with in-plane distortion.\cite{alden2013strain}
Theoretically, the electronic band structure of
the AB-BA domain boundary was recently calculated
in presence of the interlayer asymmetric potential
and it was shown that the boundary-localized states emerge
inside the asymmetry-induced energy gap. 
\cite{vaezi2013topological,zhang2013valley}

\begin{figure*}
\centerline{\epsfxsize=0.95\hsize \epsffile{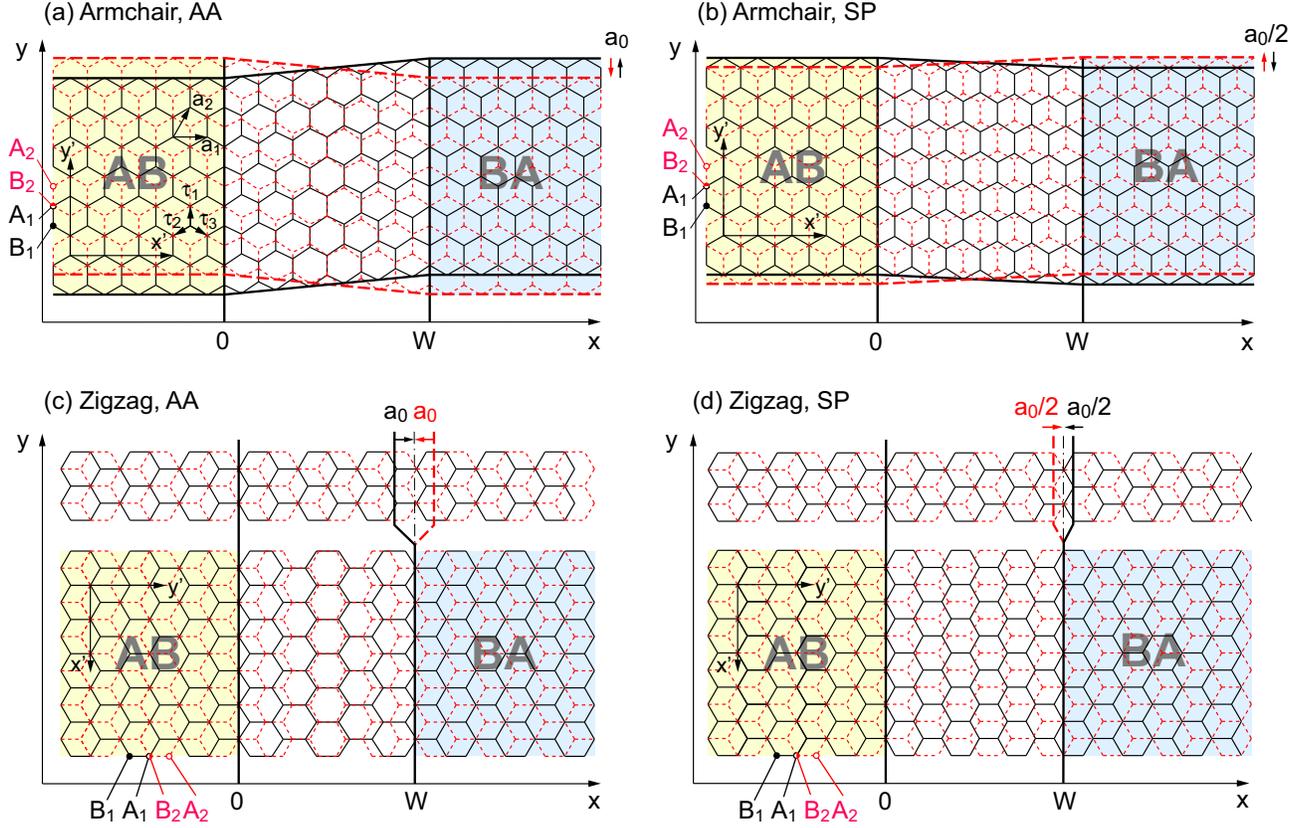}}
\caption{
Schematic structures of several types of AB-BA domain boundary:
(a) {Armchair} AA, (b) {armchair} SP, (c) {zigzag} AA, and
(d) {zigzag} SP.
In (c) and (d), a narrow strip in the upper part 
represents the uniform AB stacking bilayer before
the distortion.
}
\label{fig_atom}
\end{figure*}

In two-dimensional crystal, a domain boundary 
significantly influences the electronic transport.
There are several theoretical studies
investigating the electron transmission 
through the graphene-based domain structures,
such as the grain boundary on polycrystalline graphene 
\cite{yazyev2010electronic} and
graphene monolayer-bilayer boundary.
\cite{nakanishi2010transmission,koshino2010interface}
In this paper, we calculate the electronic transmission properties 
across AB-BA domain boundaries in bilayer graphene
with no  interlayer asymmetry.
We consider several specific  atomic configurations
illustrated in Fig.\ \ref{fig_atom}.
They are classified into {armchair} boundary [Figs.\ \ref{fig_atom} (a) and (b)]
and {zigzag} boundary [Figs.\ \ref{fig_atom} (c) and (d)]
depending on  the orientation of the boundary
relative to the honeycomb lattice. \cite{alden2013strain}
Each case is divided into AA type
[Figs.\ \ref{fig_atom} (a) and (c)]
and SP type [Figs.\ \ref{fig_atom} (b) and (d)],
depending on whether the stacking structure in the intermediate region
approximates AA stacking or SP stacking,
shown in Fig.\ \ref{fig_stacking}(c) and (d), respectively.

For each case, we calculate the transmission probability 
as a function of the incident angle and the electron energy
using the effective continuum model.
We find that the transmission strongly depends on
the boundary structure. Particularly, we show that
a low-energy electron easily passes through 
{armchair} SP [Figs.\ \ref{fig_atom} (b)],
and {zigzag} AA [Figs.\ \ref{fig_atom} (c)], 
whereas it is almost completely reflected 
in {armchair} AA [Figs.\ \ref{fig_atom} (a)],
and {zigzag} SP [Figs.\ \ref{fig_atom} (d)].
For the latter two cases, the transmission probability
sharply rises when the Fermi energy is increased to a certain level.
The {zigzag} boundaries generally have different transmission probability 
between the two different valleys, offering a possibility 
to generate the valley polarized current.
We show that those characteristic features can be
qualitatively understood by considering the local band structure
in the intermediate region.

\section{Formulation}

\subsection{Atomic structure}

We consider the AB-BA domain boundaries on bilayer graphene
defined by Figures \ref{fig_atom} (a)-(d). 
In each case, the boundary divides the system
into the AB region at left and the BA region at right.
We set the $x$ axis perpendicularly to the boundary,
and $y$ axis in parallel to the boundary.
We also define the coordinates $x'$ and $y'$ fixed to the honeycomb lattice, 
along the zigzag and armchair directions, respectively.
The relation between the two frames 
is given by $(x,y)=(x',y')$ 
for 
the {armchair} boundary [Figs.\ \ref{fig_atom} (a) and (b)],
and $(x,y)=(y',-x')$ 
for 
the {zigzag} boundary [Figs.\ \ref{fig_atom} (c) and (d)].
We define $\Vec{a}_1 = a \Vec{e}_{x'}$ and
 $\Vec{a}_2 = (1/2)a \Vec{e}_{x'} + (\sqrt{3}/2)a \Vec{e}_{y'}$ 
as the lattice vectors of graphene,
where $\Vec{e}_{x'}$ and $\Vec{e}_{y'}$ are the unit vectors 
along $x'$ and $y'$, respectively,
and $a \approx 0.246\,\mathrm{nm}$ is the lattice constant.
We also define $\GVec{\tau}_i\,(i=1,2,3)$ as the vectors from B site
to the nearest A sites as in Fig.\ \ref{fig_atom} (a).
We also use $a_0 = a/\sqrt{3}\approx 0.142$ nm
is the distance between the nearest carbon atoms in graphene. 

We define the atomic structure 
starting from uniform and infinite AB-stacked bilayer graphene.
We divide the system into the left ($x<0$), intermediate ($0<x< W$)
and right regions  ($W < x$),
where $W$ defines the width of the intermediate region.
We fix the AB stacking in the left region,
and invert AB stacking in the right region 
to BA stacking by translating layer 1 and 2 
by $\Delta y'$ and  $-\Delta y'$, respectively,
along $y'$ direction (armchair direction).
The direction of the translation is thus parallel and perpendicular
to the boundary in armchair and zigzag boundary, respectively,
resulting in shear and tensile lattice distortion, respectively.
\cite{alden2013strain}
The shift $\Delta y'$ is defined as
\begin{eqnarray}
 \Delta y' = 
\begin{cases}
a_0 & \mbox{(AA type boundary)},
\\
-a_0/2 & \mbox{(SP type boundary)}.
\end{cases}
\label{eq_delta_y}
\end{eqnarray}
For the intermediate region,
we assume the shift linearly scales in proportion to $x$.
The system is translationally symmetric along $y$ direction.

\begin{figure}
\centerline{\epsfxsize=1.\hsize \epsffile{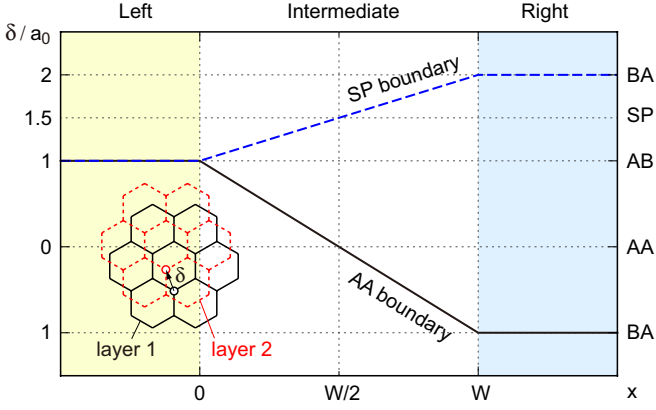}}
\caption{
The interlayer displacement $\delta(x)$ along the armchair direction  
in AA type boundary and SP type boundary.
(Inset) Bilayer graphene with a displacement $\GVec{\delta}$.
}
\label{fig_delta}
\end{figure}

The width of the intermediate region $W$ in the real system
is estimated at about 10 nm, \cite{alden2013strain}
and much larger than the atomic scale.
The local lattice structure near every single point 
then approximates a bilayer graphene with a uniform displacement.
Let us define $\GVec{\delta}$ as the interlayer displacement vector 
of layer 2 with respect to layer 1 starting from AA stacking,
as illustrated in inset of Fig.\ \ref{fig_delta}.
In the present systems, $\GVec{\delta}$ is parallel to $y'$ 
and depend only on $x$, 
so that it is written as
$\GVec{\delta}(\Vec{r}) = \delta(x) \Vec{e}_{y'}$.
$\delta(x)$ is given for the AA type boundary by,
\begin{eqnarray}
 \delta(x) =
\begin{cases}
 a_0 & (x<0) \\
 a_0(1 - 2x/W) & (0<x<W) \\
-a_0 & (W < x) 
\end{cases},
\label{eq_delta_1}
\end{eqnarray}
and for the SP type boundary,
\begin{eqnarray}
  \delta(x) =
\begin{cases}
 a_0 & (x<0) \\
 a_0(1 + x/W) & (0<x<W) \\
 2a_0 & (W < x) 
\end{cases}.
\label{eq_delta_2}
\end{eqnarray}
which are plotted in Fig.\ \ref{fig_delta}.
The AA type boundary has the AA-stacked region ($\delta = 0$)
at $x=W/2$, where the two honeycomb lattices completely overlap.
The SP boundary has the structure called SP (saddle point) stacking 
($\delta = 3a_0/2$) at $x=W/2$, 
where $A_2$ is located at the midpoint between the nearest $A_1$ sites.  
Experimentally, SP type boundary is more stable than AA type boundary
because AA stacking is energetically unfavorable. 
\cite{alden2013strain}
The total van der Waals energy of SP stacking 
is between AA stacking and AB stacking, and located at
 the saddle point on the total energy map
as a function of $\GVec{\delta}$.
\cite{popov2011commensurate,alden2013strain}

\subsection{Effective continuum model}

We derive the effective continuum Hamiltonian for AB-BA boundary
from the tight-binding model for $p_z$ atomic orbitals,
using a similar approach developed for rotationally stacked bilayer
graphene. \cite{moon2013optical}
The Hamiltonian for the tight-binding model is written as
\begin{eqnarray}
 H = -\sum_{\langle i,j\rangle}
t(\Vec{R}_i - \Vec{R}_j)
|\Vec{R}_i\rangle\langle\Vec{R}_j| + {\rm H.c.},
\label{eq_Hamiltonian_TBG}
\end{eqnarray}
where $\Vec{R}_i$ and $|\Vec{R}_i\rangle$ 
represent the lattice point and the atomic state at site $i$, respectively,
and $t(\Vec{R}_i - \Vec{R}_j)$ is
the transfer integral between the sites $i$ and $j$. 
We adopt an approximation,
\cite{nakanishi2001conductance,uryu2004electronic,trambly2010localization,slater1954simplified,moon2012energy,moon2013optical}
\begin{eqnarray}
 && -t(\Vec{d}) = 
V_{pp\pi}(d)\left[1-\left(\frac{\Vec{d}\cdot\Vec{e}_z}{d}\right)^2\right]
+ V_{pp\sigma}(d)\left(\frac{\Vec{d}\cdot\Vec{e}_z}{d}\right)^2,
\nonumber \\
&& V_{pp\pi}(d) =  V_{pp\pi}^0 
\exp \left(- \frac{d-a_0}{r_0}\right),
\nonumber \\
&& V_{pp\sigma}(d) =  V_{pp\sigma}^0 
 \exp \left(- \frac{d-d_0}{r_0}\right),
\label{eq_transfer_integral}
\end{eqnarray}
where
$V_{pp\pi}^0 \approx -2.7\,\mathrm{eV}$ 
is the transfer integral between 
the nearest-neighbor atoms of monolayer graphene,
$V_{pp\sigma}^0 \approx 0.48\,\mathrm{eV}$ is that
between vertically located atoms on the neighboring layers,
and $d_0 \approx 0.335\,\mathrm{nm}$ is the interlayer spacing.
$r_0$ is the decay length of the transfer integral,
and is chosen as $0.184 a$ so that 
the next nearest intralayer coupling becomes $0.1 V_{pp\pi}^0$.
The transfer integral for $d > 4 a_0$ is exponentially small 
and can be safely neglected.

The low-energy spectrum of the monolayer graphene
is given by effective Dirac cones centered at $K_+$ and $K_-$ points,
\cite{mcclure1956diamagnetism,divincenzo1984self,semenoff1984condensed,shon1998quantum,ando2005theory}
which are given by $\Vec{K}_\pm = \pm (2\pi/a) (2/3) \Vec{e}_{x'}$. 
We express the low-energy states of the system 
in terms of the monolayer's bases
in the vicinity of $K_\pm$ points.
It is written as $|\Psi\rangle = \sum_i \psi(\Vec{R}_i)|\Vec{R}_i\rangle$
and
\begin{eqnarray}
&& \psi(\Vec{R}_{A_l}) = 
e^{i\Vec{K}_+\cdot \Vec{R}_{A_l}} F^{K_+}_{A_l}(\Vec{R}_{A_l})
+
e^{i\Vec{K}_-\cdot \Vec{R}_{A_l}} F^{K_-}_{A_l}(\Vec{R}_{A_l})
\nonumber\\
&& \psi(\Vec{R}_{B_l}) = 
e^{i\Vec{K}_+\cdot \Vec{R}_{B_l}} F^{K_+}_{B_l}(\Vec{R}_{B_l})
+
e^{i\Vec{K}_-\cdot \Vec{R}_{B_l}} F^{K_-}_{B_l}(\Vec{R}_{B_l}),
\nonumber\\
\end{eqnarray}
where $l=1,2$ is the layer index.
Here $F^{K_\pm}_{A_l}$ and $F^{K_\pm}_{B_l}$ are 
envelope functions, which are slowly varying 
in the atomic scale.

When $W$ is much larger than the lattice constant, the interaction
between the two graphene layers is dominated by the long-wavelength
components. Then the states near the valley $K_+$ and those near $K_-$
are not hybridized 
and we may consider two valleys separately in constructing the Hamiltonian.
The effective Hamiltonian for 
$(F^{K\xi}_{A_1},F^{K\xi}_{B_1},F^{K\xi}_{A_2},F^{K\xi}_{B_2})$
is written as
\begin{eqnarray}
 {\cal H}_{\rm eff} = 
\begin{pmatrix}
H_0+V & U^\dagger \\
U & H_0-V
\end{pmatrix},
\label{eq_H_eff}
\end{eqnarray}
where
\begin{eqnarray}
&& H_0 = 
\hbar v \begin{pmatrix}
0
& \xi \hat{k}_{x'} + i \hat{k}_{y'}
\\
\xi \hat{k}_{x'} - i \hat{k}_{y'}
& 0
\end{pmatrix},
\quad 
V =
\begin{pmatrix}
0 & w^*
\\
w
& 0
\end{pmatrix},
\nonumber\\
&& U =
\begin{pmatrix}
U_{A_2A_1}
& U_{A_2B_1}
\\
U_{B_2A_1}
&
U_{B_2B_1}
\end{pmatrix}
=
\begin{pmatrix}
u(\Vec{K}_\xi,\GVec{\delta})
& u(\Vec{K}_\xi,\GVec{\delta}+\GVec{\tau}_1)
\\
 u(\Vec{K}_\xi,\GVec{\delta}-\GVec{\tau}_1)
& u(\Vec{K}_\xi,\GVec{\delta})
\end{pmatrix}.
\nonumber\\
\label{eq_submatrices}
\end{eqnarray}
$H_0$ is the effective Hamiltonian of monolayer graphene
\cite{mcclure1956diamagnetism,divincenzo1984self,semenoff1984condensed,shon1998quantum,ando2005theory},
where $\hat{\Vec{k}} = -i \partial/\partial\Vec{r}$,
$\xi = \pm$ is the valley index corresponding to $K_\pm$,
and $v$ is the band velocity of the Dirac cone,
which is given in the present tight-binding parameterization as
\cite{moon2013optical}
\begin{eqnarray}
 v \approx \frac{\sqrt{3}}{2}\frac{a}{\hbar}V_{pp\pi}^0(1-2e^{-a_0/r_0}).
\end{eqnarray}

The matrix $U$ describes the interlayer interaction 
in bilayer graphene shifted by a constant displacement $\GVec{\delta}$.
The function $u(\Vec{k},\GVec{\delta})$
is defined by \cite{moon2013optical}
\begin{eqnarray}
 u(\Vec{k},\GVec{\delta}) = 
\sum_{n_1,n_2}
- t(n_1 \Vec{a}_1 + n_2 \Vec{a}_2 + d_0\Vec{e}_z + \GVec{\delta})
\nonumber\\
\hspace{20mm}
\times
\exp\left[-i\Vec{k}\cdot(n_1 \Vec{a}_1 + n_2 \Vec{a}_2 + \GVec{\delta})
\right],
\end{eqnarray}
where $\Vec{e}_z$ is the unit vector perpendicular to graphene.
In the effective Hamiltonian, we substitute $\Vec{K}_\xi$ for $\Vec{k}$ in 
$u(\Vec{k},\GVec{\delta})$, i.e., neglect the $k$-dependent terms
in the interlayer coupling.
In AB (Bernal) stacked graphene, this simplification corresponds to
neglecting the trigonal warping and the electron hole asymmetry,
corresponding to the band parameters called $\gamma_3$ and $\gamma_4$.
\cite{mccann2006landau,koshino2009electronic}

\begin{figure*}
\centerline{\epsfxsize=0.95\hsize \epsffile{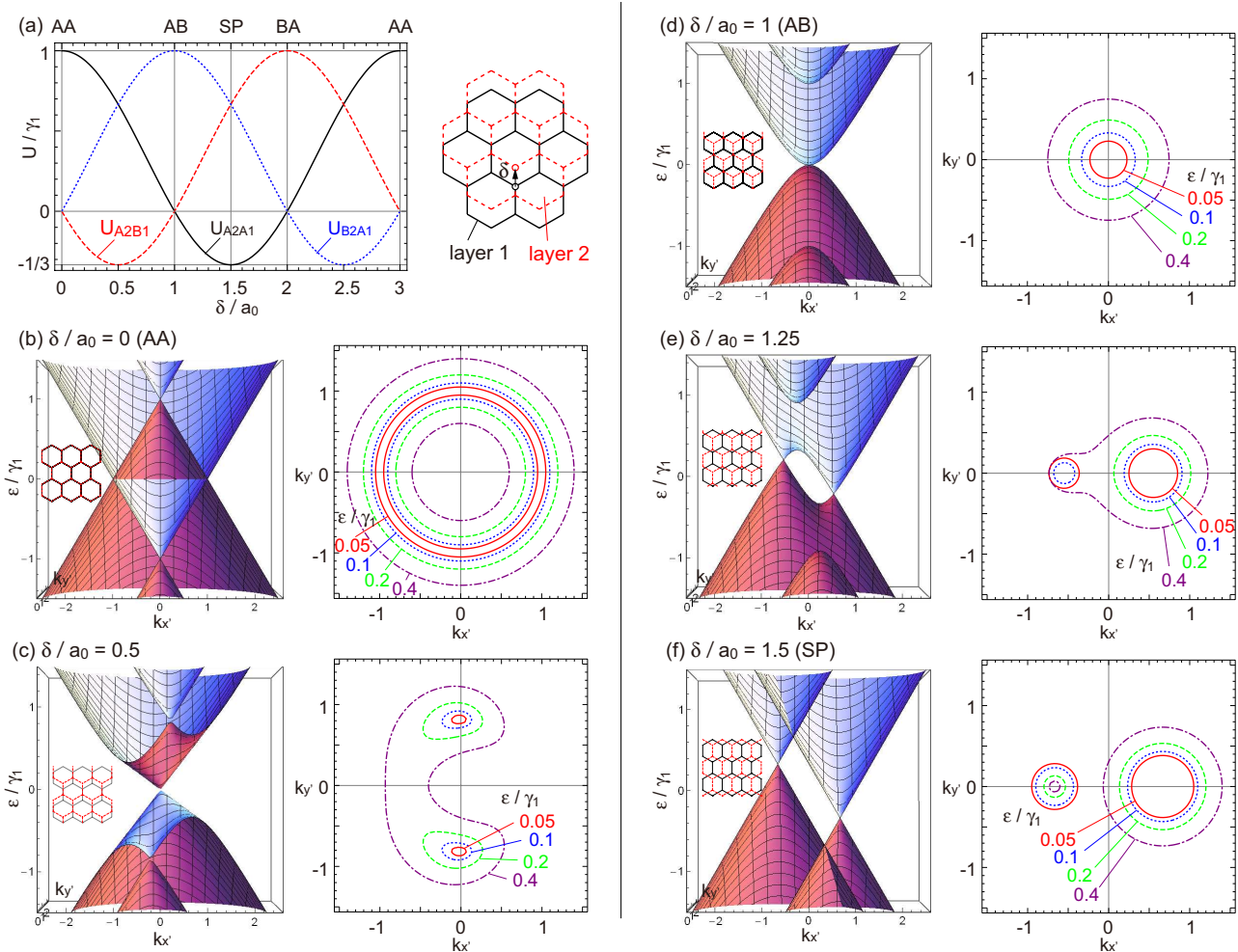}}
\caption{(a)
Interlayer matrix elements
$U_{A_2A_1}=U_{B_2B_1}$, $U_{A_2B_1}$ and $U_{B_2A_1}$
in Eq.\ (\ref{eq_u_aa}) as functions of the interlayer displacement $\delta$
along the armchair direction ($y'$).
(b)-(d) Energy band structures of bilayer graphene
with several values of $\delta$ along the armchair direction.
In each figure, the left panel 
shows the surface plot of the energy band
in $k_{y'}>0$,
and the right panel shows the contour plot
at several energies.
Energy and  wavevector are scaled in units of $\gamma_1$
and $\gamma_1/(\hbar v)$, respectively.
}
\label{fig_bilayer_shifted}
\end{figure*}

The function $u(\Vec{K}_\xi,\GVec{\delta})$ is 
smoothly varying function in $\GVec{\delta}$,
and it is approximately written in terms of a few Fourier components as
\begin{eqnarray}
&& u(\Vec{K}_\xi,\GVec{\delta})  \approx 
\frac{\gamma_1}{3} \left(
1 + e^{-i \xi\Vec{a}_1^* \cdot \mbox{\scriptsize\boldmath$\delta$}}
+ e^{-i \xi \Vec{a}_2^* \cdot \mbox{\scriptsize\boldmath$\delta$}}
\right),
\nonumber\\
&& \gamma_1 \equiv u(\Vec{K}_\xi, 0)
= t(0) -3 t(a) + 6t(\sqrt{3}a) + \cdots,
\label{eq_u_eff}
\end{eqnarray}
where $\Vec{a}^*_i$ is the reciprocal lattice vector
satisfying $\Vec{a}_i \cdot \Vec{a}^*_j = 2\pi\delta_{ij}$
and $t(x) \equiv t(x,0,d_0)$.
In the present choice of the tight-binding parameters,
we have $\gamma_1\approx 0.32{\rm eV}$.
In the following calculation, 
we scale the energy in units of $\gamma_1$,
and the wavenumber in units of $\gamma_1/(\hbar v)$,
so that the result does not depend on the values
of $v$ and $\gamma_1$.
When $\GVec{\delta} = \delta \Vec{e}_{y'}$, in particular,
Eq.\ (\ref{eq_u_eff}) becomes
\begin{eqnarray}
&& U_{A_2A_1} =  U_{B_2B_1} = 
\frac{\gamma_1}{3}
\left[
1 + 2\cos 
\, \frac{2\pi}{3}\frac{\delta}{a_0}
\right],
\nonumber\\
&& U_{A_2B_1} =
\frac{\gamma_1}{3}
\left[
1 + 2\cos 
\, \frac{2\pi}{3}\Bigl(\frac{\delta}{a_0}+1\Bigr)
\right],
\nonumber\\
&& U_{B_2A_1} =
\frac{\gamma_1}{3}
\left[
1 + 2\cos 
\, \frac{2\pi}{3}\Bigl(\frac{\delta}{a_0}-1\Bigr)
\right],
\label{eq_u_aa}
\end{eqnarray}
which are plotted against $\delta$
in Fig. \ref{fig_bilayer_shifted} (a).
The shifts $\delta=0,a_0$, $3a_0/2$ and $2a_0$ correspond to
AA, AB, SP and BA stacking, respectively, where
$(U_{A_2A_1},U_{A_2B_1},U_{B_2A_1}) =(\gamma_1,0,0)$,
$(0,0,\gamma_1)$,$\gamma_1(-1/3,2/3,2/3)$,
$(0,\gamma_1,0)$, respectively.

The matrix $V$ in Eq.\ (\ref{eq_H_eff})
describes the effect of the lattice distortion \cite{suzuura2002phonons}.
Here $w$ is given for $0<x<W$ by
\begin{eqnarray}
&& w = 
\begin{cases}
(-i \xi) \hbar v k_0 & \mbox{({armchair} boundary)}
 \\
\hbar v k_0 & \mbox{({zigzag} boundary)},
\end{cases}
\nonumber\\
&& k_0 = \frac{1}{r_0} \frac{\Delta y'}{W},
\label{eq_k0}
\end{eqnarray}
and $w=0$ otherwise,
where $\Delta y'$ is defined in Eq.\ (\ref{eq_delta_y}).
$k_0$ represents the shift of Dirac point
in the wave space.
The direction of the shift
is opposite between the two layers,
and it is always parallel to the domain boundary ($y$ direction).
As $W$ increases, the effect of $V$ becomes 
relatively unimportant compared to the interlayer interaction $U$,
since $\hbar v k_0$ in $V$ decreases as $\propto 1/W$ 
while $U$ is always of the order of $\gamma_1$.
At $W = 5\hbar v/\gamma_1 \sim 10$nm, for example,
$\hbar v k_0$ is of the order of $0.1 \gamma_1$ 
so that the electronic property is primarily determined by $U$.

\subsection{Electron transmission}

The electron transmission through the boundary can be obtained by 
using the transfer matrix. \cite{ando1991quantum,cheianov2006selective}
The effective Hamiltonian in Eq.\ (\ref{eq_H_eff})
can be written as
\begin{eqnarray}
 {\cal H}_{\rm eff}
= P \partd{}{x} + Q(x)
\end{eqnarray}
with certain matrices $P$ and $Q$,
where $Q=Q(x)$ may depend on the position $x$
and $P$ is a constant matrix.
Since the system is translationally symmetric along $y$ direction,
$\hat{k}_y = -i\partial/\partial y$ is replaced with
the quantum number $k_y$.
The Schr\"{o}dinger equation, 
$(\vare - {\cal H}_{\rm eff}) \Vec{F} = 0$,
is transformed as a one-dimensional differential equation,
\begin{eqnarray}
\partd{}{x}\Vec{F}(x) = L(x)\Vec{F}(x),
\label{eq_F_diff}
\end{eqnarray}
where $\Vec{F}=(F^{K\xi}_{A_1},F^{K\xi}_{B_1},F^{K\xi}_{A_2},F^{K\xi}_{B_2})$
is the four-component wavefunction, and
\begin{eqnarray}
L(x) = P^{-1}[\vare-Q(x)].
\end{eqnarray}

The transfer matrix is defined by
\begin{eqnarray}
 \Vec{F}(x) = T(x, x') \Vec{F}(x').
\label{eq_T_def}
\end{eqnarray}
Using Eqs.\ (\ref{eq_F_diff}) and (\ref{eq_T_def}),
the differential equation for the transfer matrix is given by
\begin{eqnarray}
 \partd{}{x}T(x, x') = L(x)T(x,x').
\label{eq_T_diff}
\end{eqnarray}
We obtain the transfer matrix  $T(W,0)$
connecting the left region and the right region
by numerically integrating Eq.\ (\ref{eq_T_diff}).


%


In the AB-stacked region in $x<0$ and
the BA-stacked region in $x>W$,
the system is uniform and $L(x)$ becomes independent of $x$.
In each region, the solution of Eq.\ (\ref{eq_F_diff}) can be written as
a linear combination of $\Vec{u}_j \exp(\lambda_j x)$,
where $\lambda_j$ and $\Vec{u}_j$ $(j=1,2,3,4)$ are
the eigen values and eigen vectors of the matrix $L$, respectively.
The four eigen values are identical in AB and BA regions and written as
\begin{eqnarray}
&&\lambda
= \pm \frac{i}{\hbar v}\sqrt{\vare^2 \pm \gamma_1 \vare - (\hbar v k_y)^2},
\end{eqnarray}
where two plus-minus signs give four possible combinations.
The corresponding eigen state is a traveling mode
when $\lambda$ is purely imaginary,
while it is an evanescent mode otherwise.
The four eigen values consist of
two right-going modes and two left-going modes,
where a right-going mode can be a
traveling mode with velocity in the positive $x$ direction,
or an evanescent mode decaying in the positive $x$ direction.

Let $\lambda_{+,i}\, (i=1,2)$ the right-going modes
and $\lambda_{-,i}\, (i=1,2)$ the left-going modes.
The wavefunction at $x=0$ is written as
\begin{eqnarray}
\Vec{F}(0) &=& 
(\Vec{u}^{\rm (L)}_{+,1},\Vec{u}^{\rm (L)}_{+,2},
\Vec{u}^{\rm (L)}_{-,1},\Vec{u}^{\rm (L)}_{-,2})
\begin{pmatrix}
 C^{\rm (L)}_{+,1}\\
C^{\rm (L)}_{+,2}\\
C^{\rm (L)}_{-,1}\\
C^{\rm (L)}_{-,2}
\end{pmatrix}
\nonumber\\
&\equiv& U^{\rm (L)} \Vec{C}^{\rm (L)},
\label{eq_F0}
\end{eqnarray}
where $\Vec{u}^{\rm (L)}_{\pm,i}$ is the eigenvector 
in the AB-region corresponding to $\lambda_{\pm,i}$,
and $C^{\rm (L)}_{\pm,i}$ is the amplitude.
Similarly, wavefunction at $x=W$ is written as
\begin{eqnarray}
\Vec{F}(W) 
&=& U^{\rm (R)} \Vec{C}^{\rm (R)},
\label{eq_FW}
\end{eqnarray}
where $U^{\rm (R)} = (\Vec{u}^{\rm (R)}_{+,1},\Vec{u}^{\rm (R)}_{+,2},\Vec{u}^{\rm (R)}_{-,1},\Vec{u}^{\rm (R)}_{-,2})$
is the eigenvectors in the BA-region.
The scripts ${\rm (L)}$ and ${\rm (R)}$ represent the left (AB) region
and the right (BA) region, respectively.
The matrices $U^{\rm (L)}$ and $U^{\rm (R)}$
are not unitary in general when they include evanescent modes.

Using Eqs.\ (\ref{eq_T_def}), (\ref{eq_F0}) and (\ref{eq_FW}),
we obtain an equation connecting the left and right wave amplitudes,
\begin{eqnarray}
&& \Vec{C}^{\rm (R)} 
= \tilde{T}\, \Vec{C}^{\rm (L)},
\nonumber\\
&& \tilde{T} = [U^{\rm (R)}]^{-1} T(W,0)  U^{\rm (L)}.
\end{eqnarray}
We write this in the form,
\begin{eqnarray}
\begin{pmatrix}
\Vec{C}^{\rm (R)}_+\\
\Vec{C}^{\rm (R)}_-
\end{pmatrix}
= 
\begin{pmatrix}
 \tilde{T}_{11} & \tilde{T}_{12} \\
 \tilde{T}_{21} & \tilde{T}_{22} 
\end{pmatrix}
\begin{pmatrix}
\Vec{C}^{\rm (L)}_+\\
\Vec{C}^{\rm (L)}_-
\end{pmatrix},
\end{eqnarray}
where $\Vec{C}_+^{\rm (R)} = (C^{\rm (R)}_{+,1},C^{\rm (R)}_{+,2})$ etc.,
and $\tilde{T}_{ij}$'s are $2 \times 2$ block matrices.
By sorting the wave amplitudes into
in-coming modes, $\Vec{C}_+^{\rm (L)}$ and $\Vec{C}_-^{\rm (R)}$, 
and out-going modes, $\Vec{C}_-^{\rm (L)}$ and $\Vec{C}_+^{\rm (R)}$,
we obtain
\begin{eqnarray}
&&\begin{pmatrix}
\Vec{C}^{\rm (L)}_-\\
\Vec{C}^{\rm (R)}_+
\end{pmatrix}
= 
S
\begin{pmatrix}
\Vec{C}^{\rm (L)}_+\\
\Vec{C}^{\rm (R)}_-
\end{pmatrix},
\nonumber\\
&&
S= 
\begin{pmatrix}
 -\tilde{T}_{22}^{-1} \tilde{T}_{21} 
& \tilde{T}_{22}^{-1} 
\\
\tilde{T}_{11}-  \tilde{T}_{12} \tilde{T}_{22}^{-1} \tilde{T}_{21}
&  \tilde{T}_{12} \tilde{T}_{22}^{-1} 
\end{pmatrix}.
\end{eqnarray}
$|S_{ij}|^2$ describes the transmission probability from the
in-coming channel $i$ to out-going channel $j$.

\begin{figure*}
\centerline{\epsfxsize=1.\hsize \epsffile{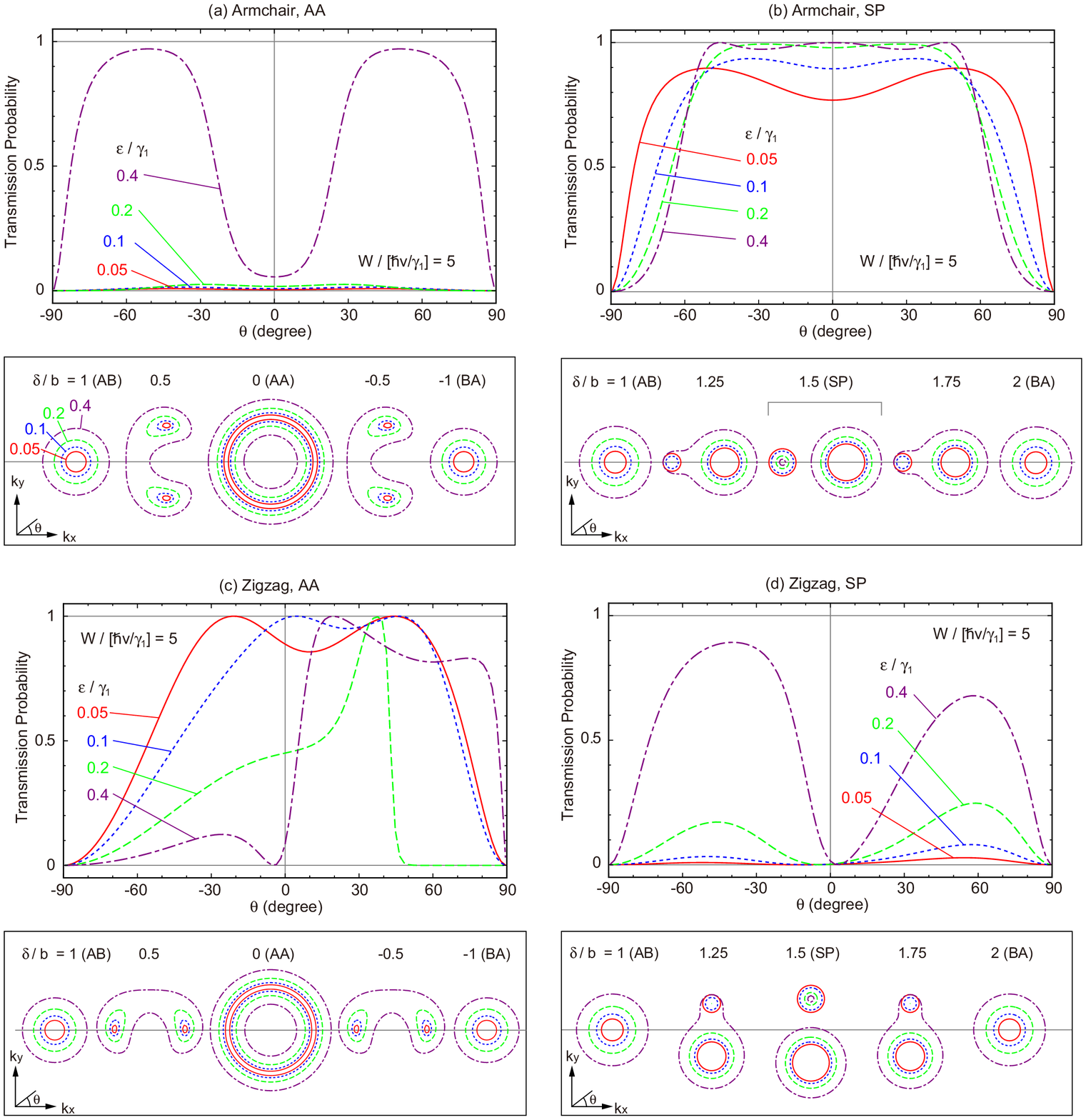}}
\caption{
Electron transmission probability $P(\theta)$ 
through the domain boundaries in Figs. \ref{fig_atom},
calculated for $K_-$ electron with several different energies.
Lower panel shows the local electronic band structure 
in the intermediate regions as a function of $\delta$.
}
\label{fig_trans}
\end{figure*}

\section{Bilayer graphene with constant displacement}
\label{sec_band_at_delta}

In the large $W$ limit
where $\GVec{\delta}$ is slowly varying in position,
the local electronic structure at every single point can be
approximately described by the electronic spectrum 
of uniform bilayer graphene with constant $\GVec{\delta}$.
The $\GVec{\delta}$-dependent band structure 
intuitively explains the transmission properties of AB-BA boundary
as shown later.
In the literature, the effect of interlayer sliding on the band structure
was studied for the cases in the vicinity of AB stacking.
\cite{mucha2011strained,son2011electronic}
For the present purposes, we consider the band structure in all the 
range of the displacement along $y'$ to cover AB, AA and SP.
Fig. \ref{fig_bilayer_shifted} (b)-(f)
show the band dispersion near $K_-$ point
for several $\GVec{\delta}$'s along AA-AB-SP line,
obtained from Eq.\ (\ref{eq_H_eff})
with constant $\GVec{\delta}$ and $V = 0$.
In each figure, the left panel 
shows the surface plot of the energy band in $k_{y'}>0$,
and the right panel the contour plot
at several energies.
The spectrum of $K_+$ are obtained by replacing $k_{x'}$
with $-k_{x'}$.

At $\delta=0$ [AA stacking, Fig. \ref{fig_bilayer_shifted} (b)]
\cite{ho2006coulomb,lee2008growth,liu2009open},
the energy spectrum consists of four bands expressed by
\begin{equation}
\vare^{\rm AA}_{s_1s_2}(\Vec{k}) = s_1 \gamma_1 +s_2 \hbar v k,
\end{equation}
where $s_1=\pm$, $s_2=\pm$, and $k=(k_x^2+k_y^2)^{1/2}$.
This represents two pairs of Dirac cones shifted by $\pm \gamma_1$
in energy, which intersect at $E=0$ with
a Fermi circle of radius $\gamma_1/(\hbar v)$.
When $\delta$ is increased from 0, the band anticrossing
occurs at the intersection, while a pair of band touching point
remains on $k_{y'}$ axis on the zero energy plane.
For $0<\delta<a_0$, the touching position is given by
\begin{equation}
 k_{x'} = 0, \quad 
k_{y'} = \pm
\frac{\gamma_1}{\hbar v}
\sqrt{
\frac{1}{3}
 + 
\frac{2}{3}\cos 
\left(\frac{2\pi}{3}\frac{\delta}{a_0}\right)
}.
\end{equation}
At $\delta=a_0$ [AB stacking, Fig. \ref{fig_bilayer_shifted} (d)], 
the split Dirac points merge at the origin,
giving the hyperbolic dispersion expressed by, 
\cite{mccann2006landau,koshino2009electronic}
\begin{equation}
\vare^{\rm AB}_{s_1s_2}(\Vec{k}) = 
s_1 \frac{\gamma_1}{2}
+
s_2 \sqrt{\left(\frac{\gamma_1}{2}\right)^2 
 + (\hbar v k)^2}.
\end{equation}
When $\delta$ exceeds 1,
the band touching point again split
but now along $k_{x'}$ axis, and also shift
in energy as shown in $\delta=1.25a_0$  [Fig.\ \ref{fig_bilayer_shifted}(e)].
At $\delta=1.5a_0$  [SP stacking, Fig.\ \ref{fig_bilayer_shifted}(f)],
the spectrum is separated into two Dirac cones
shifted in $k_{x'}$ axis the and energy axis, which are given by
\begin{equation}
\vare^{\rm SP}_{s_1s_2}(\Vec{k}) = 
s_1\frac{\gamma_1}{3} + 
s_2\sqrt{\left(\hbar v k_x+s_1 \frac{2\gamma_1}{3}\right)^2 + (\hbar v k_y)^2}.
\end{equation}
The spectrum of $\delta$ is equivalent
to that of $3 a_0 - \delta$ while the roles of $A$ and $B$
sublattices are swapped,
and thus the spectra from $\delta=1.5a_0$ to $3a_0$
is identical those from $\delta=1.5a_0$ to $0$.
The band structure is periodic in $\delta$ with period of $3a_0$.

The conduction band and the valence band are never separated by an
energy gap in any value of $\GVec{\delta}$, and this is because
both the spatial inversion symmetry and 
the time-reversal symmetry remain unbroken in  $\GVec{\delta}$. 
The robustness of the band touching point in presence of the
two symmetries was discussed for several graphene-based 
systems \cite{manes2007existence},
and it can also be explained for general systems
using the Berry phase argument, 
as shown in Appendix \ref{sec_app1}.

\section{Results and Discussion}

We calculate the electron transmission probability 
for the AB-BA boundaries of Figure \ref{fig_atom}(a)-(d).
For the width of the boundary,
we assume $W = 5\hbar v/\gamma_1 \sim 10$nm
to simulate the experimental situation \cite{alden2013strain},
and this is about ten times larger than the schematic
diagram in Fig. \ref{fig_atom}.
We consider the electron energy range $0< \vare< \gamma_1$, 
where we have a single Fermi circle
of the radius $k(\vare) = \sqrt{\vare^2 + \gamma_1 \vare}/(\hbar v)$
in the left and right regions.
We assume that an electron enters the intermediate region
from the left with energy $\vare$ and angle $\theta\, (-\pi/2<\theta<\pi/2)$,
with the initial wavevector $k(\vare)(\cos\theta,\sin\theta)$.
Since the transverse wavevector $k_y$ 
is preserved throughout the scattering process,
the electron transmits to the right region with the same angle $\theta$
with the probability $P(\theta)$,
or reflects back to the left region with the angle $\pi-\theta$
with the probability $1-P(\theta)$.

Figs.\ \ref{fig_trans}(a)-(d) show the transmission probability
$P(\theta)$ through the domain boundaries of Figs.\ \ref{fig_atom}(a)-(d),
respectively, for $K_-$ electron with several different energies.
In the {armchair} boundaries [Fig.\ \ref{fig_trans} (a),(b)], 
$P(\theta)$ is symmetric with respect to $\theta =0$
due to the $C_2$ rotation symmetry about $x$ axis (zigzag direction),
and $P(\theta)$ becomes identical in $K_+$ and $K_-$.
In the {zigzag} boundary [Fig.\ \ref{fig_trans} (c),(d)],
$P(\theta)$ is generally asymmetric, 
and $P(\theta)$ at $K_+$ is equal to $P(-\theta)$ at $K_-$.
This is because the {zigzag} boundary is symmetric
with respect to the reflection about $x$ axis (armchair direction),
which interchanges $K_+$ and $K_-$.

The transmission strongly depends on the boundary structure.
In the low energy $\vare/\gamma_1 = 0.05$, 
an electron well passes 
in {armchair} SP [Fig.\ \ref{fig_trans}(b)] and 
{zigzag} AA [Fig.\ \ref{fig_trans}(c)],
while it is almost completely reflected
in {armchair} AA [Fig.\ \ref{fig_trans}(a)] and 
{zigzag} SP [Fig.\ \ref{fig_trans}(d)].
These features can be roughly understood by considering the
local electronic band structure of fixed $\GVec{\delta}$
discussed in Sec.\ \ref{sec_band_at_delta},
which is presented in the lower 
panel in each of  Figs.\ \ref{fig_trans}(a)-(d).
In the {armchair} AA, the low-energy Fermi circle splits
along $k_y$ direction
and it prevents the electron transmission 
because there are no intermediate $k_y$  matching the initial $k_y$.
In the {armchair} SP, on the other hand, the electron well transmits 
because the Fermi circle 
splits in $k_x$ direction in this case, so that the electron 
can travel keeping the initial $k_y$.
In the {zigzag} boundaries [Fig.\ \ref{fig_trans}(c),(d)], 
the direction of the Fermi surface splitting
is rotated by $90^\circ$
so that the properties of the AA type and SP type  are 
interchanged.  

The transmission sensitively depends on the electron energy
in accordance with the change of the Fermi surface.
In the {armchair} AA, the transmission is suddenly switched on
when the energy is increased to as large as $\vare/\gamma_1 = 0.4$. 
In the band structure, correspondingly,
the Fermi circles in the intermediate region
begin to overlap with the initial Fermi circle in the AB region.
In the {zigzag} SP [Fig.\ \ref{fig_trans} (d)], 
similarly, $P(\theta)$ rises in increasing energy
as the Fermi circle overlap becomes significant.
The transmission probability remains small around
$\theta = 0$, and this corresponds to
the absence of electronic states near $k_y=0$
in the intermediate region
due to the Fermi circle splitting.
The correspondence between the transmission 
and the band structure at fixed $\GVec{\delta}$
is intuitive but only approximate, 
because the local band structure is not 
well-defined in a finite $W$, and
also the lattice distortion gives some shift of the Dirac cone.

These characteristic features of the AB-BA boundary
can be exploited to control the electronic transport. 
Particularly, the {armchair} AA [Fig.\ \ref{fig_trans}(a)] and {zigzag} SP 
[Fig.\ \ref{fig_trans} (d)] have a striking property that 
the electron transmission is almost zero near the charge neutral point,
and it sharply rises when the Fermi energy is increased to a certain
level. This suggests that the electric current through the AB-BA boundary 
can be controlled by the field effect through a single gate electrode.

\begin{figure}
\centerline{\epsfxsize=0.9\hsize \epsffile{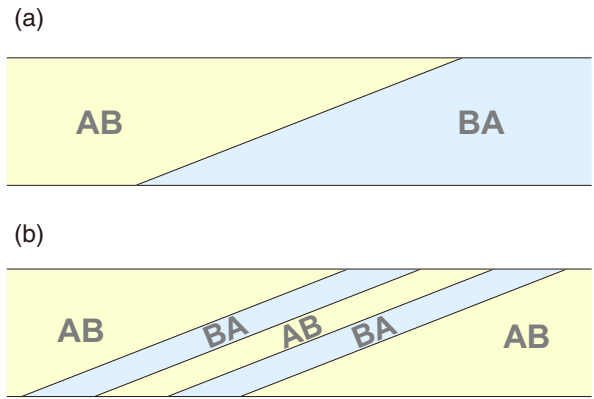}}
\caption{
Electronic channel diagonally crossing AB-BA boundary:
(a) single boundary and (b) multiple boundary case.
}
\label{fig_device}
\end{figure}

The {zigzag} boundaries generally 
have different transmission probability between $K_+$ and $K_-$
valleys, and it is possible in principle
to generate the valley polarized current.
Similar valley-selective mechanism was previously proposed 
for the graphene monolayer-bilayer junction \cite{nakanishi2010transmission},
and also for bilayer graphene with spatially modulated gate-electric field
\cite{schomerus2010helical}.
Now, the transmission probability of $K_-$ and that of $K_+$ 
through a {zigzag} boundary have opposite angle dependence,
$P(\theta)$ and $P(-\theta)$, respectively.
In the {zigzag} AA [Fig.\ \ref{fig_trans}(c)], 
$P(\theta)$ at $\vare/\gamma_1 = 0.4$ is significant only in $\theta >0$,
so that transmitted electrons are nearly polarized to $K_-$
in $\theta>0$, and to $K_+$ in $\theta <0$.
The valley polarized current can be generated 
by an electronic channel obliquely crossing the boundary,
as illustrated in Fig.\ \ref{fig_device}(a).
The polarization should be enhanced by
in multiple boundaries as in shown Fig.\ \ref{fig_device}(b).
There the transmission probability from BA to AB is identical
to that form AB to BA, because they are just related by 
interchanging the top and the bottom layers.

\begin{figure}
\centerline{\epsfxsize=0.9\hsize \epsffile{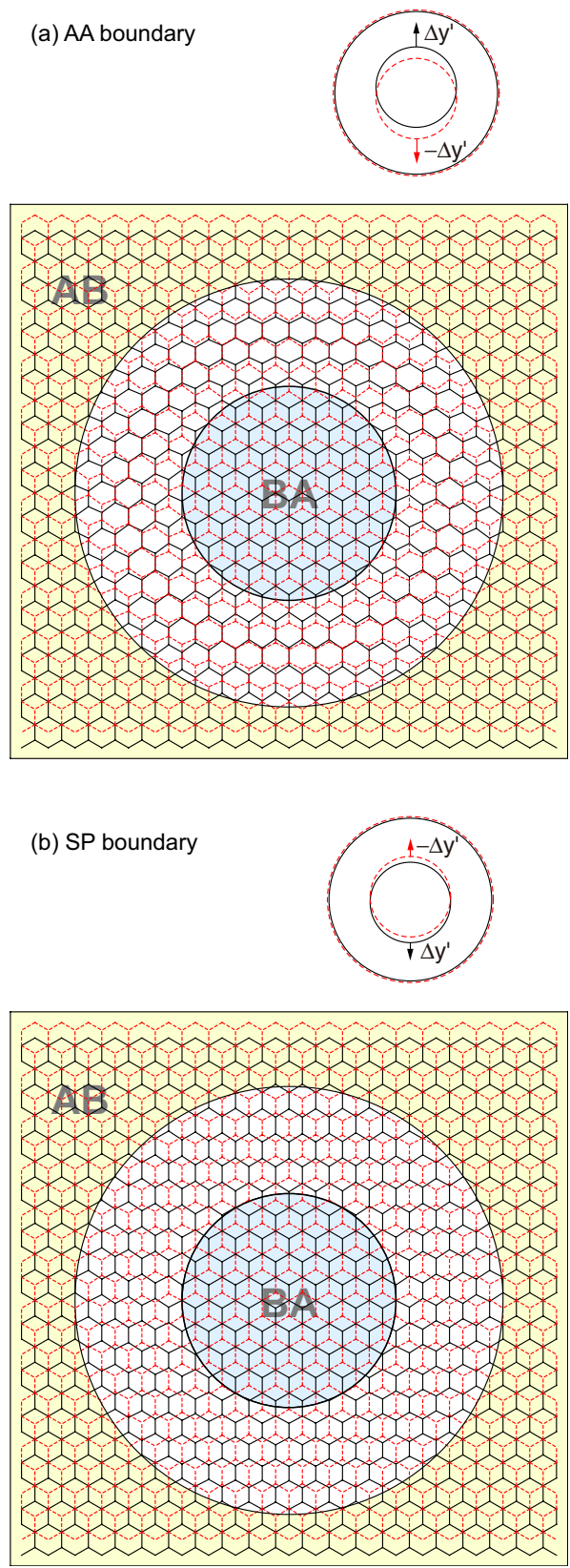}}
\caption{
AB-BA island structures
separated by (a) AA type boundary and SP type boundary,
where we translate the layer 1 and 2 inside the inner circle
by $\Delta y'$ and  $-\Delta y'$ [defined in Eq.\ (\ref{eq_delta_y})], 
respectively.
}
\label{fig_island}
\end{figure}

While we concentrated on the boundaries 
parallel to the zigzag direction or the armchair direction
in the present work,
AB-BA boundary can occur along any crystallographic direction.
Fig.\ \ref{fig_island} shows AB-BA island structures
separated by (a) AA type boundary and SP type boundary,
where we translate the layer 1 and 2 inside the inner circle
by $\Delta y'$ and  $-\Delta y'$ [defined in Eq.\ (\ref{eq_delta_y})], 
respectively,
while keeping the AB structure outside the outer circle.
In each case, we see that the armchair boundary continuously
transforms into the zigzag boundary 
as moving along the circumference. 
For a boundary along the intermediate direction
between armchair and zigzag, we can calculate the transmission 
probability in the same theoretical basis,
by rotating the crystal axis $(x',y')$ 
with respect to $(x,y)$ in the effective Hamiltonian, Eq.\ (\ref{eq_H_eff}).
We expect the transmission probability
continuously changes as a function of the boundary angle,
to interpolate the armchair and zigzag results.

\section{Conclusion}

We studied the electron transmission properties for 
AB-BA domain boundary in bilayer graphene.
Assuming several specific boundary structures,
we calculate the electron transmission probability 
as a function of the electron energy and the incident angle.
We find that the transmission strongly depends on
the boundary structure. In low-energy region,
particularly, the boundary is almost insulating 
in {armchair} AA and {zigzag} SP
while it is highly transparent in {armchair} SP and {zigzag} AA.
In insulating cases, the transmission probability 
sharply rises when the Fermi energy is increased to a certain level.
The {zigzag} boundaries generally have different transmission properties
between $K_+$ and $K_-$ valleys due to the symmetrical reason, 
and this offers a possibility to generate the valley polarized current.
The characteristic features of the electron transmission 
can be qualitatively understood by the
intermediate local band structure which continuously 
changes across the boundary.
In particular, the transport gap in {armchair} AA and {zigzag} SP 
is explained by the the wavenumber mismatch in the Fermi surface
of the intermediate region.

\section*{ACKNOWLEDGMENTS}

We thank A. W. Tsen and P. Kim for helpful discussion.
This project has been funded by 
Grant-in-Aid for Scientific Research No. 24740193
from Japan Society for the Promotion of Science (JSPS).

\appendix
\section{Robustness of Dirac points}
\label{sec_app1}

Here we show that the band touching point never disappears
in presence of the spatial inversion symmetry (SI) and 
the time-reversal symmetry (TR).
We consider a general periodic system,
and take a Bloch eigenstate
$\psi_{\Vec{k}}(\Vec{r}) = e^{i\Vec{k}\cdot\Vec{r}}u_{\Vec{k}}(\Vec{r})$.
The Berry curvature is defined as 
\begin{eqnarray}
 {\cal F}(\Vec{k}) = \nabla_{\Vec{k}} \times \Vec{A}(\Vec{k}),
\end{eqnarray}
with a vector field in $k$-space,
\begin{eqnarray}
 \Vec{A}(\Vec{k})  = -i \langle u_\Vec{k}| \nabla_\Vec{k} | u_\Vec{k} \rangle.
\end{eqnarray}
In presence of TR symmetry
and SI symmetry,
we immediately find that ${\cal F}(\Vec{k})=0$
at any non-degenerate points, because TR and SI require
${\cal F}(-\Vec{k})=-{\cal F}(\Vec{k})$
and
${\cal F}(-\Vec{k})={\cal F}(\Vec{k})$,
respectively \cite{haldane2004berry, fu2007topological}

We define the Berry phase for a closed path $C$
on the $k$-space as,
\begin{eqnarray}
 \gamma_C = \oint_C d\Vec{k}\cdot \Vec{A}(\Vec{k}).
\end{eqnarray}
By using the Stokes theorem, this is transformed as
\begin{eqnarray}
 \gamma_C = \int_S d^2\Vec{k}\cdot {\cal F}(\Vec{k}),
\end{eqnarray}
where $S$ is the $k$-space area enclosed by $C$.
In TR and SI symmetry, $\gamma_C$ can be non-zero
only when $S$ includes a band degeneracy point inside,
because otherwise ${\cal F}(\Vec{k})$ vanishes everywhere on $S$.
In a massless Dirac Hamiltonian, for example, $\gamma_C$ around
the Dirac point is $\pi$.

If we have a single degenerate point around which
$\gamma_C$ is nonzero, 
the band touching cannot be resolved in any perturbations 
which keep TR,  SI, and the original translational symmetry
(i.e., different $k$-points are not coupled).
This is because, if an infinitesimal perturbation splits the band degeneracy, 
the nonzero Berry phase around this point should immediately jump to zero
because there are no degeneracy points inside $S$ any more,
but this is obviously impossible because the
the change of the wave function on the path $C$ is also infinitesimal.
A band gap can open only when
a pair of degeneracy points having opposite Berry phases
$\gamma_C$ and $-\gamma_C$ meet and annihilate 
at a single $k$-point.



\bibliography{ab_ba_junction}

\end{document}